\documentclass[12pt]{article}
\usepackage{latexsym}
\usepackage{epsfig,amssymb,euscript}
\usepackage{amsmath}
\usepackage[pagebackref]{hyperref}

\numberwithin{equation}{section}

\def\be{\begin{equation}}
\def\ee{\end{equation}}
\def\bea{\begin{eqnarray}}
\def\eea{\end{eqnarray}}

\input{tcilatex}
\begin{document}

\date{}

%%%%%%%%%%%%%%%%%%%%%%%%%%%%%%%%%%%%%%%%%%%%%%%%%%%%%%%%%%%%%%%%%%%%%%%%

\begin{titlepage}
\begin{center}

\vskip 2.5 cm {\Large \bf

Chern-Simons theory, exactly solvable models and free fermions at finite temperature \\[3mm]}

\vskip 1.5cm
{Miguel Tierz}
 \vskip 0.8 cm

{Volen Center for Complex Systems, Brandeis University, 415 South Street, Waltham, MA 02454-9110, USA. E-mail: tierz@brandeis.edu \\
}

\end{center}

\vskip 2.25cm

\begin{abstract}
\noindent
We show that matrix models in Chern-Simons theory admit an interpretation as 1D exactly solvable models, paralleling the relationship between the Gaussian model and the Calogero model. We compute the corresponding Hamiltonians, ground-state wavefunctions and ground-state energies and point out that the models can be interpreted as quasi-1D Coulomb plasmas. We also study the relationship between Chern-Simons theory on $S^3$ and a system of N one-dimensional fermions at finite temperature with harmonic confinement. In particular we show that the Chern-Simons partition function can be described by the density matrix of the free fermions in a very particular, crystalline, configuration. For this, we both use the Brownian motion and the matrix model description of Chern-Simons theory and find several common features with c=1 theory at finite temperature. Finally, using the exactly solvable model result, we show that the finite temperature effect can be described with a specific two-body interaction term in the Hamiltonian, with 1D Coulombic behavior at large separations.

\end{abstract}

\end{titlepage}\pagestyle{plain} \setcounter{page}{1} \newcounter{bean}

\baselineskip18pt

\section{Introduction}

In addition to its original remarkable impact in topology, Chern-Simons
theory \cite{Witt} has also enjoyed a considerable interest in high-energy
physics and condensed matter physics alike (e.g. in the fractional quantum
Hall effect \cite{Fradkin}). Recall that pure Chern-Simons theory is
characterized by an action%
\begin{equation}
S_{\mathrm{CS}}(A)={\frac{k}{4\pi }}\int_{M}\mathrm{Tr}(A\wedge dA+{\frac{2}{%
3}}A\wedge A\wedge A),
\end{equation}%
with $k$ an integer number. Chern-Simons theory provides a physical approach
to three dimensional topology. In particular, it gives three-manifold
invariants and knot invariants. For example, the partition function,%
\begin{equation}
Z_{k}(M)=\int \mathcal{D}A\mathrm{e}^{iS_{\mathrm{CS}}(A)},  \label{wrt}
\end{equation}%
delivers a (quantum) topological invariant of $M$. In recent years there has
been a considerable amount of interest in pure Chern-Simons theory due to
its role in topological strings \cite{Reviews}.

Chern-Simons theory on certain manifolds is also known to have a simple
description in terms of random matrix theory \cite{Marino:2002fk,Tierz1}. In
this work, we shall employ this relationship to present a simple description
of Chern-Simons theory in terms of a one-dimensional plasma with long-range
Coulombic interactions. More precisely, we shall show that simple
one-component and two-component plasma models, with a two-body interacting
term that is a Coulomb potential of restricted dimension\footnote{%
Restricted dimension means that while the model is in $1d,$ the potential is
the one corresponding to another dimension. In our case, the surface of a $%
2d $ cylinder.}, describes Chern-Simons theory on $S^{3\text{ }}$ and on
lens spaces (to a certain extent). This happens in the following simple way:
the wavefunctional that appears in the canonical quantization of
Chern-Simons theory, can be reduced to an explicit many-body ground state
wavefunction whose Hamiltonian, which only contains one-body and two-body
potentials, is explicitly found. We also study the relationship between
Chern-Simons theory on $S^{3}$ and a system of $N$ one-dimensional fermions
at finite temperature with harmonic confinement. In particular, we show that
the partition function of Chern-Simons theory can be described by the
density matrix of the free fermions in a very particular configuration.

Let us now briefly summarize the precise relationship between Chern-Simons
theory and random matrix models, a result that we shall employ in the
following. Lawrence and Rozansky, in several works \cite{roz,lawrence},
using the exact, non-perturbative, expressions for Chern-Simons theory found
by Witten \cite{Witt}, have shown that the partition function of
Chern-Simons theory on Seifert manifolds has a simple structure and that
this partition function can be expressed entirely as a sum of local
contributions from the flat connections on the manifold. The contribution
may come from reducible and, for certain manifolds that will not be
discussed here, also irreducible flat connections. The reducible flat
connections can be expressed as an integral, instead of the complex sums
over integrable representations of the affine Lie algebra associated with
the gauge group of Chern-Simons theory \cite{Witt}. The irreducible flat
connections contributions are given by residue terms but for the two
simplest manifolds in the Seifert family, the ones that we discuss here, $%
S^{3}$ and lens spaces, only reducible flat connections contribute. In the
simplest case, that of $S^{3},$ the trivial flat connection contribution
gives the full partition function. See \cite{Marino:2002fk,Beasley} and
references therein for more details.

The connection with random matrix theory \cite{Mehta} comes from the
integral representation of the contribution of reducible flat connections.
In particular, in \cite{Marino:2002fk}, the work of Lawrence and Rozansky is
extended from $SU(2)$ to generic simply-laced gauge groups, like $U(N)$ or $%
O(2N)$ and, in addition, by specifying an orthonormal basis and expanding
the Dynkin coordinates and the positive roots of the Cartan subalgebra of
the gauge group in terms of this basis, concrete $N$-dimensional integral
expressions, reminiscent of random matrix theory are found. Let us consider
the simplest possible case, that of gauge group $U(N)$ and a $S^{3}$ manifold%
\footnote{%
The coupling constant $g_{s}$ is related with the $k$ level of Chern-Simons
theory by $g_{s}=\frac{2\pi i}{k+N}$ \cite{Marino:2002fk,Tierz1}} \cite%
{Marino:2002fk,Tierz1,Tierz2}%
\begin{equation}
Z_{\mbox{\scriptsize CS}}(S^{3})={\frac{\mathrm{e}^{-{\frac{1}{12}}%
\,N(N^{2}-1)\,g_{s}}}{N!}}\,\int \prod_{i=1}^{N}{\frac{\mbox{d}\lambda _{i}}{%
2\pi }}\,\mathrm{e}^{-|\lambda |^{2}/2g_{s}}\prod_{i<j}\left( 2\sinh {\frac{%
\lambda _{i}-\lambda _{j}}{2}}\right) ^{2}~.  \label{sinh2}
\end{equation}%
or, in matrix space \cite{Tierz1,Tierz2}%
\begin{equation}
Z_{\mbox{\scriptsize CS}}(S^{3})=\frac{\mathrm{e}^{-{\frac{1}{12}}%
\,N(7N^{2}-1)\,g_{s}}}{N!}\left( \frac{2\pi }{g_{s}}\right) ^{-N/2}\int [dM]%
\mathrm{e}^{-{\frac{1}{2g_{s}}}{\mathrm{Tr}}(\log M)^{2}}~,  \label{log2}
\end{equation}%
which are related by a simple exponential mapping \cite{Tierz1,For89}. In
this last form, it can be explicitly computed with the associated
Stieltjes-Wigert orthogonal polynomials \cite{Tierz1}, giving the known
value \cite{Witt}%
\begin{equation}
Z_{\mbox{\scriptsize CS}}(S^{3})=\frac{\mathrm{e}^{\frac{i\pi N^{2}}{4}}}{%
\left( k+N\right) ^{N/2}}\dprod\limits_{j=1}^{N-1}\left( 2\sin \frac{\pi j}{%
k+N}\right) ^{j-N}.  \label{Z}
\end{equation}%
It is well-known that the Wigner-Dyson distribution of eigenvalues $%
P_{N}^{\beta }(\left\{ \lambda _{i}\right\} )$ coincides with the
probability distribution of the $N$-particle coordinates $P_{N}^{\beta
}(\left\{ r_{i}\right\} )$ of the quantum ground state of a $1D$
Hamiltonian: the Calogero model. When there is a harmonic confinement term,
the ground state wavefunction correspond to the Gaussian random matrix
ensembles. The same property holds true for the Chern-Simons matrix model $%
\left( \ref{sinh2}\right) ,$ as we shall show and discuss in detail in the
next Section.

\section{1D exactly solvable model}

Thus, the natural question, not yet addressed, is whether one can consider:%
\begin{equation}
\Psi _{0}\left( x_{1,},...,x_{N}\right) =\sqrt{\alpha _{N}}\prod_{i=1}^{N}%
\mathrm{e}^{-\frac{x_{i}^{2}}{2g_{s}}}\prod_{i<j}\left( \sinh \frac{%
x_{i}-x_{j}}{2L}\right) ^{m},  \label{first}
\end{equation}%
so that%
\begin{equation}
Z_{N}=\langle \Psi _{0}\left\vert \Psi _{0}\right\rangle ,\text{\quad with }%
L=m=1,
\end{equation}%
with%
\begin{equation}
H\Psi _{0}=E_{0}\Psi _{0},\quad H=-\sum_{i=1}^{N}\frac{d^{2}}{dx_{i}^{2}}%
+\sum_{i=1}^{N}V(x_{i})+\sum_{i<j}W\left( x_{i}-x_{j}\right) .
\label{problem}
\end{equation}%
In principle, according to \cite{Ino}, $\left( \ref{first}\right) $ should
not correspond to an exactly solvable model satisfying $\left( \ref{problem}%
\right) $ (like Calogero model for example). A careful treatment of $\left( %
\ref{problem}\right) $ with a generic wavefunction%
\begin{equation}
\Psi _{0}\left( x_{1},...,x_{N}\right) =\prod_{i=1}^{N}\sigma \left(
x_{i}\right) \prod_{1\leq i<j\leq N}\chi \left( x_{i}-x_{j}\right) ,
\label{genericwav}
\end{equation}%
shows that there are more exactly solvable models than has been recognized
hitherto. This has been carried out in \cite{Kop}, but this work has gone
largely unnoticed. It is noteworthy that the Chern-Simons model is a
representative of this class of Hamiltonians and already a simple
computation (namely, applying the operator $-\sum \frac{d^{2}}{dx_{i}^{2}}$
in $\left( \ref{first}\right) $) shows it. This fact was already pointed out
in \cite{For} (as a counterexample to the list of exactly solvable models
provided in \cite{Ino}). Therefore, one has \cite{For}%
\begin{align}
H& =-\sum_{i=i}^{N}\frac{\partial ^{2}}{\partial x_{i}^{2}}+\frac{1}{%
g_{s}^{2}}\sum_{i=1}^{N}x_{i}^{2}+\frac{m}{g_{s}L}\sum_{i<j}(x_{i}-x_{j})%
\coth \left( \frac{x_{i}-x_{j}}{2L}\right) +\frac{m(m-1)}{2L}\sum_{i<j}\frac{%
1}{\sinh ^{2}\left( \frac{x_{i}-x_{j}}{2L}\right) },  \notag \\
E_{0}& =-\frac{m^{2}}{3}\left( \frac{1}{2L}\right) ^{2}N(N-1)(N-2)+\frac{N}{%
g_{s}}-\left( \frac{1}{2L}\right) ^{2}m^{2}N(N-1).
\end{align}%
The computation is instructive since the model has rather simple properties.
The only apparent complication are possible three-body terms. However, they
always cancel in groups of three and contribute to the energy of the ground
state with the term $\frac{1}{3}N(N-1)(N-2)$, the leading term in $E_{0}$.

Indeed, the problem $\left( \ref{problem}\right) $ with $\left( \ref%
{genericwav}\right) $ amounts to the solution of a certain functional
equation whose traditional solution (that lead to, among others, the
Calogero model) is shown in \cite{Kop} to be incomplete. More precisely, for 
$\left( \ref{problem}\right) $ with a generic wavefunction: two different
functional equations have to be solved. The first functional equation is the
one that appears when $\sigma \left( x\right) =1$:%
\begin{equation}
\varphi \left( x\right) \varphi \left( y\right) +\varphi \left( y\right)
\varphi \left( z\right) +\varphi \left( z\right) \varphi \left( x\right)
=f(x)+f(y)+f(z)\text{\quad for\quad }x+y+z=0,  \label{func1}
\end{equation}%
with $\varphi \left( x\right) =\chi ^{\prime }(x)/\chi (x).$ $\ $The second
functional equation appears because applying $\left( \ref{problem}\right) $
to $\left( \ref{genericwav}\right) $ leads to a term $\sum_{i\neq j}\frac{1}{%
2}\varphi \left( x_{i}-x_{j}\right) \left( \tau \left( x_{i}\right) -\tau
\left( x_{j}\right) \right) ,$ where $\varphi \left( x\right) $ is defined
as above and $\tau \left( x\right) =\sigma ^{\prime }\left( x\right) /\sigma
\left( x\right) ,$ that has to be expressed in terms of one-body or two-body
potentials in order to achieve exact solvability. But it turns out that what
has been usually considered is the functional equation:%
\begin{equation}
\varphi \left( x_{i}-x_{j}\right) \left( \tau \left( x_{i}\right) -\tau
\left( x_{j}\right) \right) =\lambda \left( x\right) +\lambda \left(
y\right) .  \label{funcinc}
\end{equation}%
However, as correctly pointed out in \cite{Kop}, a more general
consideration naturally leads to:%
\begin{equation}
\varphi \left( x_{i}-x_{j}\right) \left( \tau \left( x_{i}\right) -\tau
\left( x_{j}\right) \right) =\lambda \left( x\right) +\lambda \left(
y\right) +F(x-y).  \label{funccom}
\end{equation}%
The cancellations above discussed are due to the fact that for the
Chern-Simons models $f\left( x\right) $ in $\left( \ref{func1}\right) $ is a
constant. It is also a simple to see that $\lambda \left( x\right) =0$ and
that $F(x-y)=\left( x-y\right) \coth \left( x-y\right) .$

To neglect $F(x-y)$ amounts to study the possible models that can explain an
addition or modification in the confining (one-body) part of the ground
state wavefunction $\left( \ref{genericwav}\right) $ only with an addition
or modification of the confining (one-body) part of the Hamiltonian itself.
These leaves a wealth of models where such aspect of the ground state
wavefunction is described through a modification of the correlations of the
many-body problem (that is, through a two-body potential in the Hamiltonian).

Notice that the model $\left( \ref{first}\right) $ can also be interpreted
as an addition of a Gaussian one-body factor to the hyperbolic Sutherland
model, at the \ level of the wavefunction. Then, it shares $f(x)$ with this
model as this function only depends on the two-body part of the wavefunction
(namely $\varphi \left( x\right) $). In addition, as explained just above,
this modification can only be explained by the inclusion of additional
interactions at the level of the Hamiltonian, which are given in this
particular case by the term $\sum_{i<j}(x_{i}-x_{j})\coth \left( \frac{%
x_{i}-x_{j}}{2L}\right) .$ Note also that in the Coulomb gas picture, the
ground-state wavefunction is written:%
\begin{equation}
\Psi _{0}=\mathrm{e}^{-\mathcal{H}},\text{\quad }\mathcal{H=}\frac{1}{2g_{s}}%
\sum_{i=1}^{N}x_{i}^{2}-\sum_{i<j}\ln \sinh \left\vert \frac{x_{i}-x_{j}}{2L}%
\right\vert .  \label{gas}
\end{equation}%
The last interacting term is known to be the Coulomb potential between
charges in the surface of a $2D$ cylinder \cite{For}. At small distances $%
\left\vert x_{i}-x_{j}\right\vert \ll L$, the interaction term in $\left( %
\ref{gas}\right) $ behaves like the $2D$ Coulomb interaction $%
V(x_{i}-x_{j})=\ln \left\vert x_{i}-x_{j}\right\vert $, while at large
distances along the cylinder it behaves like the $1D$ Coulomb interaction $%
V(x_{i}-x_{j})=\left\vert x_{i}-x_{j}\right\vert $. The radius of the
cylinder is an additional length that conspires with $g_{s}$, the
dimensionful parameter in the quadratic confining potential, and gives rise
to the dimensionless $q$-parameter $q=\mathrm{e}^{-\frac{g_{s}}{L}}$, which
corresponds to the usual $q$-parameter in Chern-Simons theory \cite{Tierz1}
(recall the model derived in Chern-Simons theory leads to a fixed radius for
the cylinder $L=1$). A simple mapping shows this explicitly \cite%
{Tierz1,Tierz2,For}, while it allows the model to be solved exactly in terms
of $q$-orthogonal polynomials. This parameter is the responsible for the
discrete scale invariance of the model \cite{Tierz2}. In particular, the
model is known to posses Wigner solid behavior \cite{For} (see also plots in 
\cite{Tierz2}). It is worth to mention that the two-body term can also be
easily interpreted as a one-dimensional crystal potential with transverse
periodic boundary condition \footnote[1]{%
This result is immediate from basic results on crystal potentials \cite{HM}
and the elementary Fourier series $\sum_{n=1}^{\infty }\frac{1}{n}\cos
nx=-\ln \left[ 2\sin \left( \frac{\left\vert x\right\vert }{2}\right) \right]
$}. Another strong indication of crystalline behavior comes from a result in 
\cite{Tierz2}, where it was shown that the model can be discretized (with an
homogeneous lattice), without modifying its meaning in Chern-Simons theory.

The result goes beyond the gauge group and geometry above considered and,
for example, from \cite{Marino:2002fk} we can readily suggest analogous
expressions for orthogonal:%
\begin{equation}
\Psi _{0}^{SO(2N)}\left( x_{1,},...,x_{N}\right) =\prod_{i=1}^{N}\mathrm{e}%
^{-\frac{x_{i}^{2}}{2g_{s}}}\prod_{i<j}\sinh ^{m}\left( \frac{x_{i}-x_{j}}{2L%
}\right) \sinh ^{m}\left( \frac{x_{i}+x_{j}}{2L}\right) ,
\end{equation}%
and symplectic group:%
\begin{equation}
\Psi _{0}^{Sp(2N)}\left( x_{1,},...,x_{N}\right) =\prod_{i=1}^{N}\mathrm{e}%
^{-\frac{x_{i}^{2}}{2g_{s}}}\sinh \left( \frac{x_{i}}{L}\right)
\prod_{i<j}\sinh ^{m}\left( \frac{x_{i}-x_{j}}{2L}\right) \sinh ^{m}\left( 
\frac{x_{i}+x_{j}}{2L}\right) ,
\end{equation}%
again with $m=L=1$ in Chern-Simons theory. Computations are slightly more
involved, as there are more apparent three-body terms, but they cancel in
exactly the same way, leading to:%
\begin{align}
H_{SO(2N)}& =-\sum_{i=i}^{N}\frac{\partial ^{2}}{\partial x_{i}^{2}}+\frac{1%
}{g_{s}^{2}}\sum_{i=1}^{N}x_{i}^{2}+\frac{m}{g_{s}L}\sum_{i<j}\left(
(x_{i}-x_{j})\coth \left( \frac{x_{i}-x_{j}}{2L}\right) +(x_{i}+x_{j})\coth
\left( \frac{x_{i}+x_{j}}{2L}\right) \right)  \notag \\
& +\frac{m(m-1)}{2L}\sum_{i<j}\left( \frac{1}{\sinh ^{2}\left( \frac{%
x_{i}-x_{j}}{2L}\right) }+\frac{1}{\sinh ^{2}\left( \frac{x_{i}+x_{j}}{2L}%
\right) }\right) . \\
E_{0}& =-\frac{4}{3}\left( \frac{1}{2L}\right) ^{2}m^{2}N(N-1)(N-2)+\frac{N}{%
g_{s}}-2\left( \frac{1}{2L}\right) ^{2}m^{2}N(N-1).  \notag
\end{align}%
for the orthogonal, and:%
\begin{align}
H_{Sp(2N)}& =-\sum_{i=i}^{N}\frac{\partial ^{2}}{\partial x_{i}^{2}}+\frac{1%
}{g_{s}^{2}}\sum_{i=1}^{N}x_{i}^{2}+\frac{m}{g_{s}L}\sum_{i<j}\left(
(x_{i}-x_{j})\coth \left( \frac{x_{i}-x_{j}}{2L}\right) +(x_{i}+x_{j})\coth
\left( \frac{x_{i}+x_{j}}{2L}\right) \right)  \notag \\
& -\frac{2}{g_{s}}\sum_{i=1}^{N}x_{i}\coth \frac{x_{i}}{L}+\frac{m(m-1)}{2L}%
\sum_{i<j}\left( \frac{1}{\sinh ^{2}\left( \frac{x_{i}-x_{j}}{2L}\right) }+%
\frac{1}{\sinh ^{2}\left( \frac{x_{i}+x_{j}}{2L}\right) }\right) . \\
E_{0}& =-\frac{4}{3}\left( \frac{1}{2L}\right) ^{2}m^{2}(N+1)N(N-1)+N\left( 
\frac{1}{g_{s}}-1\right) -2\left( \frac{1}{2L}\right) ^{2}m^{2}N(N-1). 
\notag
\end{align}%
for the symplectic case. In contrast to $\left( \ref{first}\right) $, these
models have not been discussed elsewhere. As mentioned above, invariants
other than the partition function can be considered. It is well-known that
knot invariants can be obtained with \cite{Witt}%
\begin{equation}
Z_{k}(M)=\int \mathcal{D}AW_{R}^{\mathcal{K}}\left( A\right) \mathrm{e}^{iS_{%
\mathrm{CS}}(A)},
\end{equation}%
where $W_{R}^{\mathcal{K}}\left( A\right) $ is the Wilson loop operator,
which is the trace of the holonomy around the knot. It is possible to
generalize the results of \cite{Marino:2002fk} to other knot invariants \cite%
{Dolivet:2006ii}. For example, the case of torus knots is conjectured to be
given by \cite{Marino}%
\begin{equation}
W_{R}^{(P,Q)}={C}_{N}(P,Q)\int \prod_{i=1}^{N}dx_{i}S_{\lambda _{\ell }}(%
\mathrm{e}^{x_{i}})\mathrm{e}^{-\frac{x_{i}^{2}}{2g_{s}}}\prod_{i<j}\left(
2\sinh {\frac{x_{i}-x_{j}}{2P}}\right) \left( 2\sinh {\frac{x_{i}-x_{j}}{2Q}}%
\right) ,  \label{int}
\end{equation}%
where $\Lambda $ is the highest weight corresponding to the representation
labelled by $\lambda $, and shifted by $\rho $ and $S_{\lambda }(x_{i})$ are
Schur polynomials associated to the partition $\lambda $.

The case $P=Q=1$ corresponds to the unknot. Then, the above integral reduces
to the expression of the partition function of Chern-Simons theory with an
insertion of a Schur polynomial $S_{\lambda }$. In this case, using the
explicit expression for the Schur polynomial:%
\begin{equation}
S_{\lambda }(\mathrm{e}^{u})={\frac{\sum_{w\in \mathcal{W}}\epsilon (w)%
\mathrm{e}^{\Lambda \cdot w(u)}}{\sum_{w\in \mathcal{W}}\epsilon (w)\mathrm{e%
}^{\rho \cdot w(u)}}},
\end{equation}%
together with Weyl's denominator formula,%
\begin{equation}
\sum_{w\in \mathcal{W}}\epsilon (w)\mathrm{e}^{w(\rho )}=\prod_{\alpha
>0}2\sinh {\frac{\alpha }{2}},
\end{equation}%
the above integral $\left( \ref{int}\right) $ reduces to a Gaussian, and can
be performed exactly. One gets in fact:%
\begin{equation}
W_{\Lambda }(\mathrm{unknot})={\frac{S_{\rho \Lambda }}{S_{\rho \rho }}}
\end{equation}%
and $W_{\Lambda }(\mathrm{unknot})$ is $\dim _{q}\lambda $ , the quantum
dimension of $\Lambda $. Thus, employing the ground-state wavefunctions one
can write for the quantum dimensions%
\begin{equation}
\dim _{q}\lambda =\langle \Psi _{0}\left\vert S_{\lambda }\left( \mathrm{e}%
^{x}\right) \right\vert \Psi _{0}\rangle .  \label{qdim}
\end{equation}%
In \cite{Dolivet:2006ii}, we give an explicit computation, using a mixture
of combinatorial and orthogonal polynomials methods, of the integral that
appears in the r.h.s. of $\left( \ref{qdim}\right) .$

A more detailed study is clearly required in order to see if we can extend
the description to excited states. For example, to see if $\left( \ref{qdim}%
\right) $ can be understood as an scalar product of excited states of the
many-body Hamiltonian. For the moment, let us just show that other, more
complex models, that appear in the literature, for example when one
considers lens spaces, exhibit the same properties. We just present the
result when the geometry is $S^{3}/\mathbb{Z}_{2}$, a case of interest in
topological string theory (see \cite{Reviews} for a review). If we consider
the corresponding (two-)matrix model \cite{Aganagic:2002wv}, then the wave
function is%
\begin{eqnarray*}
\Psi _{0}^{S^{3}/\mathbb{Z}_{2}}\left(
x_{1},...,x_{N_{1}},y_{1,}...,y_{N_{2}}\right) &=&\tprod\limits_{i=1}^{N_{1}}%
\mathrm{e}^{-\frac{x_{i}^{2}}{2g_{s}}}\tprod\limits_{j=1}^{N_{2}}\mathrm{e}%
^{-\frac{y_{j}^{2}}{2g_{s}}}\prod_{1\leq i<j\leq N_{1}}\sinh \left( \frac{%
x_{i}-x_{j}}{2}\right) \times \\
&&\prod_{1\leq i<j\leq N_{2}}\sinh \left( \frac{y_{i}-y_{j}}{2}\right)
\prod_{i,j}\cosh \left( \frac{x_{i}-y_{j}}{2}\right) .
\end{eqnarray*}%
Remarkably enough, explicit computation shows that all the three-body terms
cancel as before, leading to%
\begin{align}
H_{S^{3}/\mathbb{Z}_{2}}& =-\sum_{i=i}^{N_{1}}\frac{\partial ^{2}}{\partial
x_{i}^{2}}-\sum_{i=i}^{N_{2}}\frac{\partial ^{2}}{\partial y_{i}^{2}}+\frac{1%
}{g_{s}^{2}}\left(
\sum_{i=1}^{N_{1}}x_{i}^{2}+\sum_{i=1}^{N_{2}}y_{i}^{2}\right) +\frac{1}{%
g_{s}L}\sum_{1\leq i<j\leq N_{1}}(x_{i}-x_{j})\coth \left( x_{i}-x_{j}\right)
\notag \\
& +\frac{1}{g_{s}L}\sum_{1\leq i<j\leq N_{2}}(y_{i}-y_{j})\coth \left(
y_{i}-yj\right) +\frac{1}{g_{s}L}\sum_{i,j}\left( x_{i}-y_{j}\right) \tanh
\left( x_{i}-y_{j}\right) .
\end{align}%
Since the corresponding matrix model is now a two-matrix model we have found
the correspondence with a two-component plasma rather than the one-component
plasmas above discussed.

To summarize this Section, we have presented one and two-component Coulomb
plasmas that are also solutions of the $\mathrm{1D}$\ Hamiltonian problem $%
\left( \ref{problem}\right) $ (like Calogero models, for example), and that
provide a rather simple many-body description of Chern-Simons theory, for
several gauge groups and geometries, mimicking its ability to deliver
quantum topological invariants.

\section{Free fermions at finite temperature}

%\tableofcontents

As we have explained above, Chern-Simons theory on certain manifolds has a
simple description in terms of random matrix theory \cite%
{Marino:2002fk,Tierz1}. In this Section, we shall employ this relationship
to point out a connection between Chern-Simons theory and free fermions at
finite temperature. In the rich interplay between supersymmetric topological
strings, $c=1$ non-critical bosonic strings, IIB superstrings and
Chern-Simons theory, this is possibly the less well-known one: the
relationship between Chern-Simons theory and $c=1$ theory. This was already
mentioned in \cite{Dijkgraaf:2003xk} as the weakest link in the manifold
connections between the above mentioned theories and models. An early study
of the relationship between Chern-Simons theory and free fermions appeared
long ago in \cite{Douglas}. Most of the considerations in this Section
amount to rearrangements of what is previously known in the literature, but
this rearrangement may be illuminating as some connections have not been
recognized hitherto.

Before proceeding, recall that the structure of the $\mathcal{N}=2$
topological strings parallels that of bosonic topological strings. In \cite%
{Ghoshal:1995wm} it was shown that critical $\mathcal{N}=2$ topological
strings are mapped into $c=1$ non-critical strings.\ More precisely, they
showed that the $c=1$ non-critical string corresponding to a CFT on a circle
at the self-dual radius is equivalent to a topological $\mathcal{N=}2$
theory at the conifold. A particular consequence of this is that the genus
expansion of the free energy of the $c=1$ string at self-dual radius
coincides with the same expansion for the free energy when $N=\infty$ of $%
SU(N)$\ Chern-Simons on $S^{3}$ \cite{Ghoshal:1995wm}. In general, the
relationship between $c=1$ theory and Chern-Simons theory, as the one we
have just mentioned involves topological matrix models (see \cite%
{Mukhi:2003sz} for a review), while we shall discuss a free fermion picture
which involves the double-scaled matrix models of $2D$ quantum gravity.
Regarding the relationship between these double-scaled models of $c=1$
matrix quantum mechanics and Chern-Simons theory, it is only known that the
free energy, when the target space is a circle, is very similar to the
nonperturbative part of Chern-Simons theory on $S^{3}$\ and gauge group $%
U(N) $ (see \cite{Nakayama:2004vk} for a recent review and discussions).

The matrix model description reviewed in the Introduction is in turn
intimately related to a Brownian motion description of Chern-Simons theory 
\cite{deHaro:2004id}, which is based on non-intersecting random walkers. The
particular process that is related to Chern-Simons theory is that of $N$
vicious walkers on a line. Walkers are vicious \cite{fisher} if they
annihilate each other when they meet. If we denote their coordinates by $%
\lambda _{i}$, $i=1,\ldots ,N$, they satisfy $\lambda _{1}>\lambda
_{2}>\ldots >\lambda _{N}$. Alternatively, this process can be regarded as
motion of a single particle in the fundamental Weyl chamber of $U(N)$. The
particle starts moving at position $\mu _{i}$ satisfying $\mu _{1}>\mu
_{2}>\ldots >\mu _{N}$, and is required to stay within the Weyl chamber. The
process stops when the particle hits one of the walls. One then computes the
probability of going from an initial position $\mu _{i}$ to a final position 
$\lambda _{i}$ staying always within the chamber. This is given by \cite%
{fisher}:%
\begin{equation}
p_{t,N}(\lambda ,\mu )={\frac{1}{(2\pi t)^{N/2}}}\,\mathrm{e}^{-{\frac{%
|\lambda |^{2}+|\mu |^{2}}{2t}}}\,\det |\mathrm{e}^{\lambda _{i}\mu
_{j}/t}|_{1\leq i<j\leq N}~.  \label{two}
\end{equation}

The quickest way to make contact with Chern-Simons theory is to evaluate
this amplitude in a very specific case: we take the same initial and final
boundary conditions, i.e. $\mu =\lambda $, and an equal spacing condition,
that is, $\lambda _{0j}-\lambda _{0,j+1}=a$, where $a$ is the initial and
final spacing between two neighboring movers. We compute the probability of
a reunion after time $t$. Notice that, since the $\lambda $'s also label
highest weights of irreducible representations of $U(N)$, this boundary
condition is labeled by the Weyl vector for a suitable choice of the overall
scale. Now a straightforward computation yields:%
\begin{equation}
p_{t,N}(\lambda _{0},\lambda _{0})={\frac{1}{(2\pi t)^{N/2}}}%
\,\prod_{k=1}^{N}(1-\mathrm{e}^{-ka^{2}/t})^{N-k}~.
\end{equation}%
If one chooses units where $a^{2}=1$ and identifies -$\frac{1}{t}=g_{s}=%
\frac{2\pi i}{k+N},$ then this is the Chern-Simons $S^{3}$ $U(N)$ partition
function. What about the explicit relationship between $\left( \ref{two}%
\right) $ and $\left( \ref{sinh2}\right) $ ? In \cite{deHaro:2004id}, we
showed that the matrix model expression for the partition function of
Chern-Simons on $S^{3}$ $\left( \ref{sinh2}\right) $ corresponds to the
extensivity property of probabilities:%
\begin{equation}
p_{t+t^{\prime },r}(\rho ,\rho )=\int [\mbox{d}\lambda ]\,p_{t,r}(\rho
,\lambda )\,p_{t^{\prime },r}(\lambda ,\rho ),  \label{additivity}
\end{equation}%
where the range of integration is the same as in the matrix model. Now, the
elementary but important step is to realize that $\left( \ref{two}\right) $
is also the density matrix description of a system of free fermions in one
dimension at finite temperature $T$ and with harmonic confinement%
\begin{align}
\rho \left( y,x\right) & =C_{N}\det_{ij}\left\langle y_{i}\right\vert \exp
(-\beta \left( \frac{\widehat{p}^{2}}{2m}+\frac{1}{2}m\omega ^{2}\widehat{x}%
^{2}\right) )\left\vert x_{j}\right\rangle  \label{density} \\
& =C_{N}\det_{ij}\left( \exp \left( -\frac{m\omega }{2\hbar }\coth \left(
\beta \hbar \omega \right) (y_{i}^{2}+x_{j}^{2})+\frac{m\omega }{\hbar }%
\frac{y_{i}x_{j}}{\sinh \left( \beta \hbar \omega \right) }\right) \right) 
\notag
\end{align}

Indeed, this quantity is well-known from the early works that studied $c=1$
theory at finite temperature \cite{Gross:1990ub,Gross:1990md,Boulatov:1991xz}
(that is, with target space a circle instead of the line). Recall that
string theory on a circle can be studied by a chain of matrices model that,
in turn, was solved using a transfer matrix approach \cite%
{Gross:1990ub,Gross:1990md} and the transfer matrix operator is precisely $%
\left( \ref{density}\right) $, which is also the propagator of the
upside-down harmonic oscillator \cite{Gross:1990ub,Boulatov:1991xz}. In this
correspondence we have that the string coupling constant is directly
proportional to the temperature $g_{s}\sim T\sim 1/t.$ Note that this
relationship is opposite to the one that appears in the crystal melting
picture \cite{crystal}.

Thus, the underlying system is in principle the same: $N$ one-dimensional
free fermions harmonically confined at finite temperature. What is the
difference between the Chern-Simons theory and string theory on a circle ?
The Chern-Simons partition function is given by the overlap between
identical initial and final states after temporal evolution in Euclidean time%
\footnote{%
Euclidean time $t$ corresponds to the inverse of the string coupling
constant $g_{s}$ and to the radius of the cylinder in the Brownian motion
description (see the Conclusions).} 
\begin{equation}
Z_{\mbox{\scriptsize CS}}(S^{3})=p_{t,N}(\lambda _{0},\lambda _{0})=\rho
\left( x_{0},x_{0}\right) ,
\end{equation}%
where $x_{0}$ denotes a very particular position configuration with the
fermions equispaced with unit distance. On the other hand, the partition
function of string theory on a circle%
\begin{equation}
Z=\mathrm{Tre}^{-\beta RH}=\sum_{a}\int \left\langle \phi _{a}\right\vert
\exp (-\beta H)\left\vert \phi _{a}\right\rangle =\sum_{a}\rho \left(
x_{a},x_{a}\right) ,
\end{equation}%
implies the sum over all states.

In what follows, we shall discuss the free fermion at finite $\mathrm{T}$
behavior directly from the one matrix model formulation $\left( \ref{sinh2}%
\right) ,\left( \ref{log2}\right) $ and from the point of view of exactly
solvable models. Before proceeding, let us point out that the probability
distribution $\left( \ref{two}\right) $ has also been considered in a very
different setting in random matrix theory. More precisely, in the study of
critical statistics \cite{critical}. The term critical statistics denotes a
statistics of eigenvalues that interpolates between the Wigner-Dyson
statistics \cite{Mehta} (typical of Gaussian matrix models and the one that
is relevant in the metallic phase of a disordered system) and uncorrelated
Poissonian statistics (insulating regime). Several matrix models in
condensed matter physics have been introduced that incorporate this critical
statistic, in an attempt to describe the metal to insulator transition in
disordered systems. One of these models is the Moshe-Neuberger-Shapiro
model, defined by \cite{Moshe:1994gc}%
\begin{equation}
\mathcal{P}_{U}(H)\mathrm{d}^{N^{2}}H=C_{N}\exp \left( -\mathrm{Tr}H^{2}-b%
\mathrm{Tr}\left( \left[ U,H\right] \left[ U,H\right] ^{\dagger }\right)
\right) \mathrm{d}^{N^{2}}H,  \label{MNS}
\end{equation}%
where $U$ is an unitary matrix and $U=V^{\dagger }DV$ with $D_{ij}=\delta
_{ij}\mathrm{e}^{i\theta _{i}}.$ The idea of \cite{Moshe:1994gc} is that $U$
defines a preferred basis with the $b$-dependent term, trying to align the
Hermitian matrix $H$ with $U$ and hence leading to a preference for the
basis $V.$ Even after averaging over $U(N)$ with the invariant $U(N)$ Haar
measure (which restores unitary invariance of the model) one obtains
different statistics from the traditional Wigner-Dyson statistics. Note that
the $b$-dependent term is of the same type as the one that naturally appears
in matrix quantum mechanics \cite{Gross:1990md}%
\begin{equation}
\mathrm{Tr}\left( \overset{\centerdot }{\Phi }\right) ^{2}=\mathrm{Tr}\left[ 
\overset{\centerdot }{\Lambda }^{2}+\left[ \Lambda ,A\right] \left[ \Lambda
,A\right] \right] ,
\end{equation}%
with the matrix $\Phi (t)=\Omega \left( t\right) \Lambda \left( t\right)
\Omega \left( t\right) $ with $\Omega \ $unitary and $A(t)=$ $\overset{%
\centerdot }{\Omega }\Omega ^{\dagger }$ a pure gauge field. This average
over $U(N)$, as with the $c=1$ matrix model \cite{Gross:1990md}, is done
using the Harish-Chandra-Izykson-Zuber integral, obtaining \cite%
{Moshe:1994gc}%
\begin{equation}
P(x_{1},...,x_{N})=\det_{ij}\left( \exp \left( -\left( b+\frac{1}{2}\right)
\left( x_{i}^{2}+x_{j}^{2}\right) +2bx_{i}x_{j}\right) \right)
\end{equation}%
which is again $\left( \ref{two}\right) $ or, equivalently, $\left( \ref%
{density}\right) $ only that already in the diagonal representation. So, it
is a particular case of $\left( \ref{two}\right) $ or $\left( \ref{density}%
\right) $, but the one relevant in Chern-Simons theory. The connection of
this last expression with free fermions at finite temperature is also made
in \cite{Moshe:1994gc}. It is also worth to mention that one recurrent topic
in many of the works that discuss the different random matrix models of
critical statistics is to try to explain the fact that seemingly different
matrix models, in particular $\left( \ref{MNS}\right) $ and $\left( \ref%
{log2}\right) $ lead to the same statistics (their two-point correlation
kernel, that we shall discuss below, coincide). The explanation, as we have
seen above, is essentially contained in \cite{deHaro:2004id}, as the matrix
model representation comes from expressing the density probability as the
composition of two processes, using an intermediate step to finally arrive
at the departure point $\left( \ref{additivity}\right) .$

Now, as stated above, the connection between the Chern-Simons matrix models
and free fermions at finite temperature can also be seen from the properties
of the corresponding one-matrix model $\left( \ref{sinh2}\right) ,\left( \ref%
{log2}\right) $. The property is actually valid for $q$-deformed matrix
model in general and for the Chern-Simons matrix model in particular. Recall
that free fermions in one dimension are characterized by a first-order order
coherence function:%
\begin{equation}
g\left( \overrightarrow{r}\right) =\left\langle \widehat{\psi }^{\intercal
}(r)\widehat{\psi }(0)\right\rangle =\frac{\sin \left( k_{0}r\right) }{\pi r}%
,  \label{coherence}
\end{equation}%
which is the two-point correlation of an Hermitian matrix model:%
\begin{equation}
K(x,y)=\frac{\sin (\pi \left( x-y\right) )}{\left( x-y\right) },
\end{equation}%
the famous sine kernel, describing the bulk correlations of a Hermitian
Gaussian matrix model \cite{Mehta}. In the finite temperature case, it was
found in \cite{1976} (see \cite{Castin} for a modern and direct treatment)
that $\left( \ref{coherence}\right) $ is replaced by:%
\begin{equation}
g\left( \overrightarrow{r}\right) =\frac{\delta k_{0}\sin k_{0}r}{\sinh
\left( \pi r\delta k_{0}\right) },  \label{finiteT}
\end{equation}%
where $\delta k_{0}$ is the momentum width where the Fermi surface is no
longer infinitely sharp, but has a smooth variation with an energy width of $%
k_{B}T$. More precisely, we have:%
\begin{equation}
\delta k_{0}=\frac{mk_{B}T}{\hslash ^{2}k_{0}}.
\end{equation}%
It turns out that the two-point kernel of Hermitian random matrix ensembles
with a $q$-deformed weight (for $\mathrm{e}^{-\pi ^{2}/a}\ll 1$) \cite%
{critical} 
\begin{equation}
K(x,y)=\frac{a\sin (\pi \left( x-y\right) )}{\sinh \left( a\pi \left(
x-y\right) \right) },\text{ where }a:=\frac{1}{2}\log \frac{1}{q}=\frac{g_{s}%
}{2},
\end{equation}%
which is $\left( \ref{finiteT}\right) $ with the $a$-parameter denoting the
momentum width $\delta k_{0}$ and hence $g_{s}\sim T$ as above. The zero
temperature limit implies $g_{s}\rightarrow 0$ and consequently $%
q\rightarrow 1$. This is the limit where the model goes to the Gaussian
(GUE) universality class, which has a free fermion at zero temperature
description, so we have a consistent result. The opposite limit $%
q\rightarrow 0$ (temperature tending to infinite) is less studied and will
be discussed elsewhere (see the Conclusions). Note that finite temperature
leads to a decrease of the correlations as the two-point function now
decreases exponentially. The model $\left( \ref{sinh2}\right) $ is of course
more fluctuating than an ordinary Wigner-Dyson ensemble (a Gaussian model
for example), with a higher level of repulsion between eigenvalues, and
hence there is less correlation between particles, as they prefer to stay
further away from each other. This can be made more precise at the level of
the Hamiltonian, by recalling the results of the first Section. That is to
say, it can be explained through additional interactions responsible for
this temperature effect, that shows up as an strengthened repulsion at the
level of the eigenvalue distribution. Let us show and briefly discuss here
the explicit form of this additional interaction, that lead to the decrease
of correlations typical of finite temperature effects on a system of
fermions. In the Chern-Simons case we have seen that%
\begin{equation}
H=-\sum_{i=i}^{N}\frac{\partial ^{2}}{\partial x_{i}^{2}}+\frac{1}{g_{s}^{2}}%
\sum_{i=1}^{N}x_{i}^{2}+\frac{m}{g_{s}}\sum_{i<j}(x_{i}-x_{j})\coth \left( 
\frac{x_{i}-x_{j}}{2}\right) +\frac{m(m-1)}{2}\sum_{i<j}\frac{1}{\sinh
^{2}\left( \frac{x_{i}-x_{j}}{2}\right) },
\end{equation}%
is the Hamiltonian whose ground-state is the Chern-Simons eigenvalue
distribution $\left( \ref{sinh2}\right) $%
\begin{equation}
H\Psi _{0}=\mathrm{E}_{\mathrm{0}}\Psi _{0},\quad \Psi
_{0}(x_{1},...,x_{N})=\prod_{i=1}^{N}\,\mathrm{e}^{-x_{i}^{2}/2g_{s}}%
\prod_{i<j}\left( 2\sinh {\frac{x_{i}-x_{j}}{2}}\right) ^{m}.\ 
\end{equation}%
Consider the case of a Hermitian matrix model $m=1$ (the probability
distribution of the matrix model is interpreted as the square of the
wavefunction)$,$ that leads to the free fermion case in the
Gaussian/Calogero model case. We see that even in this case a two-body
correlation term survives%
\begin{equation}
V(x_{i}-x_{j})=\frac{1}{g_{s}}\sum_{i<j}(x_{i}-x_{j})\coth \left( \frac{%
x_{i}-x_{j}}{2}\right) .  \label{interact2}
\end{equation}%
This is the term responsible for the departure with the usual free fermion
at zero temperature behavior. Note how close is this interaction to be a
Coulomb potential, as for a large separation between particles we have $%
V(x-x)=\left\vert x_{i}-x_{j}\right\vert $. Thus, it is the Coulomb
potential in one-dimension, but modified at small separations with
essentially a constant term (instead of going to $0,$ as the Coulomb
potential does). This result makes precise the observation in \cite{CM},
where it is argued that the temperature effect can be described by
additional interaction terms. Note also that this last result leads to an
equivalent interpretation in terms of $\mathrm{N}$ fermions in one dimension
with harmonic confinement, and two-body interactions described by $\left( %
\ref{interact2}\right) .$

\section{Conclusions and Outlook}

We have seen that Chern-Simons partition functions can be obtained in two
different ways:

\begin{itemize}
\item[(i)] As the norm of $\left\vert \Psi_{0}\right\vert ^{2}$ as is
typical of Chern-Simons theory $Z_{\mbox{\scriptsize CS}}(S^{3})=\langle\Psi
_{0}\left\vert \Psi_{0}\right\rangle ,$ but with the $N$-body wavefunctions
instead of wavefunctionals.

\item[(ii)] Specifying a very particular configuration where the fermions
are equispaced as described in Section 3.
\end{itemize}

In the first Section, we explained that the model $\left( \ref{first}\right) 
$ describes a quasi-one dimensional Coulomb plasma, since it lives in $1D$
but the repulsion between eigenvalues term ($V(x_{i}-x_{j})=\ln \sinh \left(
x_{i}-x_{j}\right) $ in the Coulomb gas picture) is the Coulomb potential
between two charges on the surface of a cylinder. This coincides with the
non-intersecting Brownian motion description, as the returning condition for
the $N$ walkers on the line implies that, in Euclidean time (as is the case
of the matrix model description -$\frac{1}{t}=g_{s}$), we have $N$
non-intersecting Brownian motions on a cylinder, with the radial direction
of the cylinder being the Euclidean time. Thus, the returning condition,
essential to obtain the Chern-Simons partition function as we have seen,
leads to the finite temperature result because it makes the Euclidean time
to be compactified on a circle. This cylindrical geometry also emerges in
the $D$-brane derivation of the matrix model \cite{Aganagic:2002wv}.

Regarding the exactly solvable model, not only the connection with
Chern-Simons theory is new but also the model has been poorly studied in
itself. Hence, there are several open problems. For example, integrability
of the model is an open problem and to find a Lax pair, provided the model
is integrable, is a possible concrete task. It may also be of interest to
study excited states of the model (something well-known for the Calogero
model, and that involves the appearance of Jack polynomials, that comprise
Schur functions), and see if they are related with states of the type $%
\left( \ref{int}\right) $ with $P=Q=1$ or, equivalently, with $\left( \ref%
{qdim}\right) $. In general, we believe that the appearance in Chern-Simons
theory of exactly solvable models that are so related to the celebrated
Calogero and Sutherland models (see the introduction and for details on
their relationship) is an interesting result that deserves further attention.

Besides topological strings, where Chern-Simons matrix models have already
proved to be of interest, the models involved may play a direct role in
condensed-matter physics as well. For example, it is immediate to identify $%
\left( \ref{first}\right) $ as the model that appears in a one-dimensional
representation of a Laughlin state on a cylinder. An even more direct
application has to do with the fact that other exactly solvable models with $%
F(x-y)\neq 0$ may be also of relevance in the study of Bose gases, as a
generalization of the celebrated Lieb-Liniger model \cite{Tierzprep} and
that these models are directly related to the Chern-Simons matrix model
discussed here, in the limit where the cylindrical geometry above discussed
is very thin.

Other physical applications besides topological strings, quantum Hall effect
and Bose gas models, may have to do with the fact that quantum topological
invariants play a role in the characterization of topological order. In the
recent works \cite{top}, we find a connection between topological order and
quantum dimensions. In general, a rich interplay between quantum
computation, Chern-Simons theory and condensed matter physics has recently
emerged. It would be certainly interesting if the many-body description of
Chern-Simons theory and its topological invariants presented here could be
employed in the context of topological quantum computation \cite{tqc1}.

\setcounter{secnumdepth}{-1}

\subparagraph{Acknowledgements.}

The author is grateful to Eduardo Fradkin, Jaume Gomis, Sebastian de Haro
and Marcos Mari\~{n}o for comments and correspondence.

%
%\cite{Witten:1988hf}

\end{document}